\documentclass[preprint,amsmath,amssymb,aps,showkeys,showpacs]{revtex4}
\usepackage[english]{babel}
\usepackage{graphicx}
\usepackage{graphics}
\usepackage{amsmath}
\usepackage{dcolumn}
\usepackage{amssymb}
\usepackage{bm}

\begin{document}
\title{On the Majorana fermion subject to a linear confinement }
\author{R. F. Ribeiro} 
\email{rfreire@fisica.ufpb.br}
\affiliation{Departamento de F\'isica, Universidade Federal da Para\'iba, Caixa Postal 5008, 58051-900, Jo\~ao Pessoa-PB, Brazil.}

\author{K. Bakke}
\email{kbakke@fisica.ufpb.br}
\affiliation{Departamento de F\'isica, Universidade Federal da Para\'iba, Caixa Postal 5008, 58051-900, Jo\~ao Pessoa-PB, Brazil.}

\begin{abstract}

We analyse the linear confinement of a Majorana fermion in $\left(1+1\right)$-dimensions. We show that the Dirac equation can be solved analytically. Besides, we show that the spectrum of energy is discrete, however, the energy levels are not equally spaced.

\end{abstract}
\keywords{Majorana fermions, linear scalar potential, relativistic wave equations, analytical solutions}
\pacs{03.65.Pm, 03.65.Ge, 31.30.J-, 71.15.Rf}

\maketitle

\section{Introduction}

Majorana fermions have an interesting characteristic: each particle is its own anti-particle \cite{maj,maj3,maj4}. This characteristic of the Majorana fermions has allowed a great advances in studies of the quantum computation by exploring the topological quantum computation \cite{maj6}. Majorana fermions have also been widely explored in condensed matter physics with studies of Berry's phase \cite{maj2,maj5}, semiconductor devices \cite{maj7}, topological insulators \cite{maj8} and quantum wires \cite{maj9}. Electromagnetic properties of the Majorana fermions have been analysed in Refs. \cite{maj10,maj11} and a pedagogical discussion about Dirac, Majorana and Weyl fermions has been made in Ref. \cite{maj12}. In quantum field theory, Majorana fermions have been discussed in neutrino physics \cite{maj13}, supersymmetry \cite{maj14} and dark matter \cite{maj15}.

The aim of this work is to analyse the linear confinement of a Majorana fermion \cite{maj} in $\left(1+1\right)$-dimensions. From the studies of confinement of quarks \cite{linear1,linear4d,linear4a,linear4b,linear4c} to atomic and molecular physics \cite{linear3a,linear3b,linear3c,linear3d,linear3e,linear3f,fb}, the linear scalar potential has a great interest. By following Ref. \cite{greiner}, we can introduce a scalar potential by modifying the mass term of the relativistic equations (Dirac or Klein-Gordon equations as examples), i.e., we rewrite the mass term as $m\rightarrow m+S\left(x\right)$, where $S\left(x\right)$ is the scalar potential. It is worth pointing out that, by modifying the mass term of either Dirac or Klein-Gordon equations, we build a relativistic analogue of the position-dependent mass systems \cite{pdm,pdm1,pdm2}. Effects associated with the linear confinement have also been investigated in different relativistic quantum system \cite{linear2,linear2a,linear2b,linear2c,linear2d,linear2e,linear2f,vb,vb2} and in the presence of topological defects in the spacetime \cite{vb3,eug,eug2}. Hence, we search for relativistic bound state solutions to a Majorana fermion subject to a linear scalar potential.

The structure of this paper is: in section II, we discuss the behaviour of a Majorana fermion subject to a linear scalar potential in $\left(1+1\right)$-dimensions, where we show that the Dirac equation can be solved analytically; in section III, we present our conclusions.

\section{Linear confinement}

In this section, we investigate the interaction of a Majorana fermion with a linear scalar potential in $\left(1+1\right)$-dimensions. Since we wish to work in $\left(1+1\right)$-dimensions, then, the line element of the Minkowski spacetime is written in the form (with the units $c=\hbar=1$):
\begin{eqnarray}
ds^{2}=-dt^{2}+dx^{2}.
\label{1.1}
\end{eqnarray} 
Thereby, the Dirac equation that describes a Majorana fermion subject to a linear scalar potential is given by
\begin{eqnarray}
i\gamma^{\mu}\partial_{\mu}\psi-\left(m+b\,x\right)\psi=0,
\label{1.2}
\end{eqnarray} 
where $b$ is a constant that characterizes the scalar potential and the $\gamma^{\mu}$ matrices are defined in the Majorana representation:
\begin{eqnarray}
\gamma^{0}=\sigma^{2};\,\,\,\,\,\gamma^{1}=i\,\sigma^{3},
\label{1.3a}
\end{eqnarray}
with $\vec{\sigma}=\left(\sigma^{1},\,\sigma^{2},\,\sigma^{3}\right)$ as being the Pauli matrices. It is worth pointing out that the Majorana representation of  the $\gamma^{\mu}$ matrices given in Eq. (\ref{1.3a}) has the properties $\left(\gamma^{\mu}\right)^{*}=-\gamma^{\mu}$, $\left(\gamma^{\mu}\right)^{\dag}=\gamma^{0}\,\gamma^{\mu}\,\gamma^{0}$ and obeys the rules of Clifford algebra $\left\{\gamma^{\mu},\,\gamma^{\nu}\right\}=-2\,\eta^{\mu\nu}$, where $\eta^{\mu\nu}=\mathrm{diag}\left(-\,+\right)$ is the Minkowski tensor \cite{maj3,maj4,greiner}. In this way, $\psi$ describes a Majorana fermion, and the Dirac equation (\ref{1.2}) becomes
\begin{eqnarray}
i\,\frac{\partial\psi}{\partial t}=m\,\sigma^{2}\,\psi+i\,\sigma^{1}\,\frac{\partial\psi}{\partial x}+m\,x\,\sigma^{2}\,\psi
\label{1.3}
\end{eqnarray}
A solution to Eq. (\ref{1.3}) is given in the form
\begin{eqnarray}
\psi\left(t,\,x\right)=e^{-i\mathcal{E}t}\,\left(
\begin{array}{c}
\phi\left(x\right)\\
\chi\left(x\right)\\	
\end{array}\right).
\label{1.4}
\end{eqnarray}
By substituting the solution (\ref{1.4}) into the Dirac equation (\ref{1.3}) we obtain two coupled equation for $\phi$ and $\chi$, where the first coupled equation is
\begin{eqnarray}
\mathcal{E}\phi=i\left(\frac{d}{dx}-m-bx\right)\chi,
\label{1.5}
\end{eqnarray}
while the second coupled equation is given by
\begin{eqnarray}
\mathcal{E}\chi=i\left(\frac{d}{dx}+m+bx\right)\phi.
\label{1.6}
\end{eqnarray}
By eliminating $\chi$ in Eqs. (\ref{1.5}) and (\ref{1.6}), we obtain the following second order differential equation:
\begin{eqnarray}
\phi''-b^{2}\left(\frac{m}{b}+x\right)^{2}\phi+\left(\mathcal{E}^{2}+b\right)\phi=0.
\label{1.7}
\end{eqnarray}

Let us define $x_{0}=\frac{m}{b}$, and thus, perform a change of variables given by $r=\sqrt{b}\left(x+x_{0}\right)$, then, Eq. (\ref{1.7}) becomes
\begin{eqnarray}
\frac{d^{2}\phi}{dr^{2}}-r^{2}\,\phi+\frac{\beta}{b}\,\phi=0.
\label{1.8}
\end{eqnarray} 

Observe that the second order differential equation (\ref{1.8}) has singular points as $r\rightarrow0$ and $\left|r\right|\rightarrow\infty$. By analysing the asymptotic behaviour of the possible solutions to Eq. (\ref{1.8}) as $r\rightarrow0$ and $\left|r\right|\rightarrow\infty$, then, the solution to Eq. (\ref{1.8}) is given by
\begin{eqnarray}
\phi\left(r\right)=e^{-\frac{r^{2}}{2}}\,\,F\left(r\right),
\label{1.10}
\end{eqnarray}
where $F\left(r\right)$ is an unknown function. Further, by substituting Eq. (\ref{1.10}) into Eq. (\ref{1.8}), we obtain the following equation for $F\left(r\right)$:
\begin{eqnarray}
F''-2r\,F'+\left(\frac{\beta}{b}-1\right)F=0.
\label{1.11}
\end{eqnarray}

Our next step is to search for polynomial solutions to Eq. (\ref{1.11}), then, we write the function $F\left(r\right)$ as a power series expansion around the origin, i.e., $F\left(r\right)=\sum_{k=0}^{\infty}a_{k}\,r^{k}$ \cite{griff,arf}. Thereby, by substituting this series into Eq. (\ref{1.11}), we obtain a recurrence relation given by:
\begin{eqnarray}
a_{k+2}=\frac{\left(2k+1-\frac{\beta}{b}\right)}{\left(k+2\right)\left(k+1\right)}\,a_{k}.
\label{1.12}
\end{eqnarray}
From Eq. (\ref{1.12}), hence, the series terminates, i.e., it becomes a polynomial of degree $n$, by imposing that \cite{griff,arf}:
\begin{eqnarray}
\frac{\beta}{b}=2n+1,
\label{1.15}
\end{eqnarray}
where $n=0,1,2,3,\ldots$. From the condition established in Eq. (\ref{1.15}), we obtain 
\begin{eqnarray}
\mathcal{E}_{n}=\pm\sqrt{2\,n\,b}.
\label{1.16}
\end{eqnarray}

Hence, Eq. (\ref{1.16}) gives us the relativistic energy levels of a Majorana fermion subject to a linear confinement in $\left(1+1\right)$-dimensions. Observe that the mass term of the Dirac (Majorana) equation is absent in Eq. (\ref{1.16}), however, it yields the change in the $x$-origin given by $x_{0}=\frac{m}{b}$. Therefore, there is no gap between $\mathcal{E}_{n}^{+}=+\sqrt{2\,n\,b}$ and $\mathcal{E}_{n}^{-}=-\sqrt{2\,n\,b}$, and the energy of the ground state is null. Besides, the energy levels are not equally spaced, i.e., $\mathcal{E}_{n+1}-\mathcal{E}_{n}=\sqrt{2\left(n+1\right)b}-\sqrt{2\,n\,b}$, for any value of $n$.

\section{conclusions}

We have investigated the linear confinement of a Majorana fermion in $\left(1+1\right)$-dimensions and show that the corresponding relativistic wave equation can be solved analytically. We have found that the spectrum of energy is discrete, however, the energy levels are not equally spaced. We have also found that the mass term is absent from the relativistic energy levels, which means that there is no gap between the positive and negative energies. The contribution of the mass term is yielding the change in the $x$-origin given by $x_{0}=\frac{m}{b}$. 

Since we have dealt with analytical solutions of a relativistic wave equation, the present work can be in the interest of studies of position-dependent mass systems \cite{pdm,pdm1,pdm2}, semiconductor devices \cite{maj7}, topological insulators \cite{maj8}, effects of topological defects on quantum systems \cite{put,prl} and quantum wires \cite{maj9}. An interesting point of discussion about this discrete spectrum of energy given in Eq. (\ref{1.16}) is the possibility of building coherent states \cite{coh,coh2,coh3,coh4} and displaced Fock states \cite{disp,lbf}.

\acknowledgments{We would like to thank C. Furtado for interesting discussions. We also thank CNPq (Conselho Nacional de Desenvolvimento Cient\'ifico e Tecnol\'ogico - Brazil) for financial support.}

\end{document}